\documentstyle[emulateapj,onecolfloat]{article}

\def\etal{{\rm et~al.\ }}
\def\hmpc{\;h^{-1}{\rm Mpc}}

\def\hmpccc{\;h^{3}{\rm Mpc}^{-3}}
\def\hkpc{h^{-1}{\rm kpc}}

\def\msun{{\rm h^{-1} M_{\odot}}}
\def\shear{\langle \gamma^{2} \rangle}
\newcommand{\PSbox}[3]{\mbox{\rule{0in}{#3}\includegraphics{#1}\hspace{#2}}}
\begin{document}
 
\twocolumn[
 
\title{Weak lensing surveys and the intrinsic correlation of
galaxy ellipticities}
 
\author{
Rupert A.C.~Croft and Christopher A.~Metzler
}  
\affil{Harvard-Smithsonian Center for Astrophysics, Cambridge, MA 02138}
\authoremail{rcroft@cfa.harvard.edu}

\begin{abstract}
We explore the possibility that an intrinsic correlation between 
galaxy ellipticities arising during the galaxy formation process
may account for part of the shear signal recently reported by several 
groups engaged in weak gravitational lensing surveys.
Using high resolution N-body simulations we measure the projected
 ellipticities 
of dark matter halos and their correlations as a function of pair
separation. 
With this simplifying, but not necessarily
realistic assumption (halo shapes as a proxy for galaxy
shapes), we find a positive detection of correlations
up to  scales of at least $20 \hmpc$ (limited by the box size). The signal
is not strongly affected by variations in 
 the halo finding technique, or by the resolution of the simulations
(over the range tested).
We translate our three dimensional results into angular measurements 
of ellipticity correlation functions and shear variance which can be 
directly compared to observational results. We also make simulated 
angular surveys by projecting our simulation boxes onto the plane of the 
sky and applying a radial selection function. Measurements from these
catalogs are consistent with the analytic projection of the statistics.
Interestingly, the shear variance we measure is a small, but 
 not entirely 
negligible fraction (from $\sim10-20 \%$, depending on the angular scale)
of that seen by the observational groups, and the ellipticity
correlation functions approximately
 mimic the functional form expected to be caused
by weak lensing. The amplitude of these projected quantities
depends strongly on the width in redshift of the galaxy distribution.
If in the future photometric redshifts are used to  pick out a screen
of background galaxies with a small redshift width, then the intrinsic
correlation may become comparable to the weak lensing signal.
Although we are dealing with simulated dark matter halos, we 
might expect there to be a similar sort of signal when real galaxies are used.
This could be checked fruitfully using a nearby sample with known redshifts.
\end{abstract}
 
]

\section{Introduction}

The large--scale mass distribution in the Universe is expected, through
gravitational lensing, to imprint itself on the pattern of
ellipticities measured from  background galaxies (e.g., Blandford \etal 1991;
Miralda-Escud\'{e} 1991).  The angular correlations of such
ellipticities, or the variance in ellipticities
averaged in angular cells (amongst other statistics) can be compared to
expectations for different cosmological models, and in principle can
discriminate between them (e.g., Blandford \etal 1991; Miralda-Escud\'{e} 1991;
Kaiser 1992; Jain \& Seljak 1997).
Since lensing is induced by the foreground
mass distribution only, it provides the most direct method of studying
the structure of mass in the Universe on large scales.  Accordingly, detecting
the shear signal induced by large scale structure has been the subject
of a great deal of observational effort. 
 Recently, four separate
groups have reported detections of this cosmic shear, at levels
comparable to that expected from currently popular models of structure
formation (Van Waerbeke \etal 2000 [hereafter VW];
 Bacon, Refregier \& Ellis 2000;
Wittman \etal 2000; Kaiser, Wilson \& Luppino 2000 [hereafter KWL]).
 
In attributing the observed correlations of ellipticities, or the shear
variance, to large--scale structure, an important assumption is that the
sample of background galaxies used contains no intrinsic correlation of
ellipticities.  If such a correlation were present in the sample of
lensed sources, it could be attributed to lensing, and may enhance
the detected signal.  The argument for discounting this possibility is
that a pair of galaxies separated by
a small distance on the sky are nonetheless on average separated by a
large distance along the line of sight.  If a particular pair of galaxies
are separated by a large distance, there is no good theoretical reason
to expect their ellipticites to be intrinsicly correlated.
 
However, the angular correlation of ellipticites predicted to be due to 
 lensing
is quite small --- on the order of $10^{-4}$ out to scales of several
arcminutes.  To detect this correlation against the random variations
of galaxy ellipticities, observers take deep images and use large
samples of background galaxies, perhaps $10^{5}$ per square degree.
Some of these galaxies can be expected to be close not only in projection,
but in real space as well.  To what degree should we expect correlations
in the actual ellipticities of nearby galaxies, and how much would the
projection of such correlations add to any observed lensing signal?
 
That such intrinsic correlations in galaxy ellipticities may exist is
not implausible.  If elongation by local tidal fields contributed
significantly to galaxy ellipticities, then nearby galaxies could
be expected to sample the tidal field in the same fashion, producing
similar elongations.  Alternately, if some elongation originating
from a galaxy's last merger were to survive for a time comparable
to the characteristic merger timescale for its environment, then
one might expect galaxies to be preferentially aligned along the
local large--scale structure, and thus similarly to each other.
On larger scales, such elongation in cosmic structures appears
to be present.  For instance,
the effects of large scale structure on the shapes
of galaxy clusters have been the subject of much study.
Cosmological N--body simulations have suggested that clusters tend
to be oriented towards neighboring clusters or in directions defined
by adjoining filaments and the merging subclusters which drain along
them. (Dekel, West \& Aarseth 1984; West, Dekel \& Oemler 1989;
West, Villumsen \& Dekel 1991;
van Haarlem \& van de Weygaert 1993; Splinter \etal 1997;
de Theije, van Kampen \& Slijkhuis 1998).  Observations have typically
indicated the presence of such alignments, either towards nearby
clusters (Binggeli 1982; Flin 1987; West 1989a,b;
Rhee, van Haarlem \& Katgert 1992; Plionis 1994)
or towards nearby large--scale structure
in the galaxy distribution (Argyres \etal 1986; Lambas, Groth \&
Peebles 1988); although not all studies support the presence of
such alignments (Struble \& Peebles 1985; Ulmer, McMillan \& Kowalski 1989;
Fong, Stevenson \& Shanks 1990).  The existence of correlations
in the alignment of large--scale structures appears quite
possible; perhaps similar intrinsic correlations in alignment
exist on galactic scales.
 
Theoretical expectations for the degree of correlation of intrinsic
ellipticities can in principle be derived.  The local gravitational
shear can be expected to either align the intrinsic angular momentum
of nearby galaxies (Lee \& Pen 2000), or to similarly deform
neighboring, non--rotating galaxies through tidal distortion
(Ciotti \& Dutta 1994).  Thus, the statistics of the local tidal
field can be related to the statistics of galaxy angular momenta
(Catelan \& Theuns 1996a,b; Catelan \& Theuns 1997; Sugerman,
Summers \& Kamionkowski 1999); and therefore
to the intrinsic correlations in galaxy ellipticities (Coutts 1996;
Lee \& Pen 2000;
Catelan \etal 2000, in preparation;  Mackey and White 2000, in preparation).
 
There have been numerous attempts to detect intrinsic correlations
in galaxy alignments using low redshift samples; the picture painted
by this work is unclear, as we can see from the following sample.
Flin (1988) considered a sample of 118 galaxies in the
Perseus supercluster and found that the spin axes of these galaxies
 were aligned with the supercluster plane.
Muriel \& Lambas (1992) reported a correlation of alignments seen
with spirals taken from the ESO catalog and analyzed in three
dimensions; when only projected data was considered, the correlation
was no longer present.
Garrido \etal (1993) analyzed a sample covering a large area
of sky in the northern hemisphere and claimed to find no evidence
for correlations in alignment except within the Coma supercluster.
Han, Gould \& Sackett (1995) examined the spins of 60 galaxies
in the Ursa Major filament and found no evidence for any alignment
of spins.  Cabanela \& Aldering (1998) considered galaxy shapes
extracted from a survey of Perseus--Pisces conducted using an
automated plate scanner; statistically significant and color
dependent correlations of galaxy ellipticities were found.
On the other hand, Cabanela \& Dickey (1999) used HI observations to
determine the spins of 54 galaxies in the Perseus--Pisces supercluster;
and found no evidence for preferential alignments of spin vectors.
At this time evidence favors an
orientation alignment between cD galaxies and the major axis of
their parent cluster; but the presence or absence of any other galaxy shape
correlations remains undetermined.
 
In this paper, we use Nbody simulations to make theoretical
predictions for the correlation of intrinsic galaxy ellipticities.
For simplicity, we work with directly with the 
projected ellipticities of the simulated dark matter
halos, without making assuming any model for the way galaxies form 
within them. The significance of our results will therefore 
be entirely dependent on whether galaxy ellipticities
behave significantly differently from
their halos, a problem we leave to future gasdynamical simulations.

The layout of the paper is as follows. 
In \S2, we describe the N--body dataset used, our halo
catalog, and our measurement of projected ellipticities from the halos.
We measure the three dimensional correlation functions
of projected ellipticities in \S3, and in 
 \S4 we decribe the construction of simulated surveys from 
our halo catalogs, with a geometry designed to mimic
weak lensing observations.
In \S5, we project the three dimensional correlation functions into angular 
statistics, including the shear variance. We also compute these 
 angular measures directly
from our simulated surveys (as a consistency check).
We compare our results to current observational data in \S6, before
discussing and summarizing our results in \S7.

\section{Simulated halos}

\subsection{Nbody simulations}
Our requirements are that the simulation volume be large enough to 
capture the large scale tidal field that may cause 
correlations to arise between halo shapes, while at the same
time having enough mass resolution to follow the formation
of galaxy sized halos with a reasonable number of particles. 
We use outputs from Nbody simulations run by the Virgo Consortium (see e.g.,
the author list of Jenkins \etal 1998 for Virgo members) and which they have
generously made public. The simulations are part of a set of different models,
although we only use one here,  the currently favoured cosmological
constant-dominated cold dark matter ($\Lambda$CDM) model. The parameters of
this model are as follows, $\Omega_{m}=0.3$ (the matter density at $z=0$),
$\Omega_{\Lambda}=0.7$ (the contribution of $\Lambda$ to the critical
 density at z=0), $\Gamma=0.21$, which is the shape parameter of the initial
power spectrum. The 
 normalization was set so that $\sigma_{8}=0.9$, (the rms matter fluctuations
in $8 \hmpc$ spheres extrapolated to $z=0$).

\begin{figure*}[t]
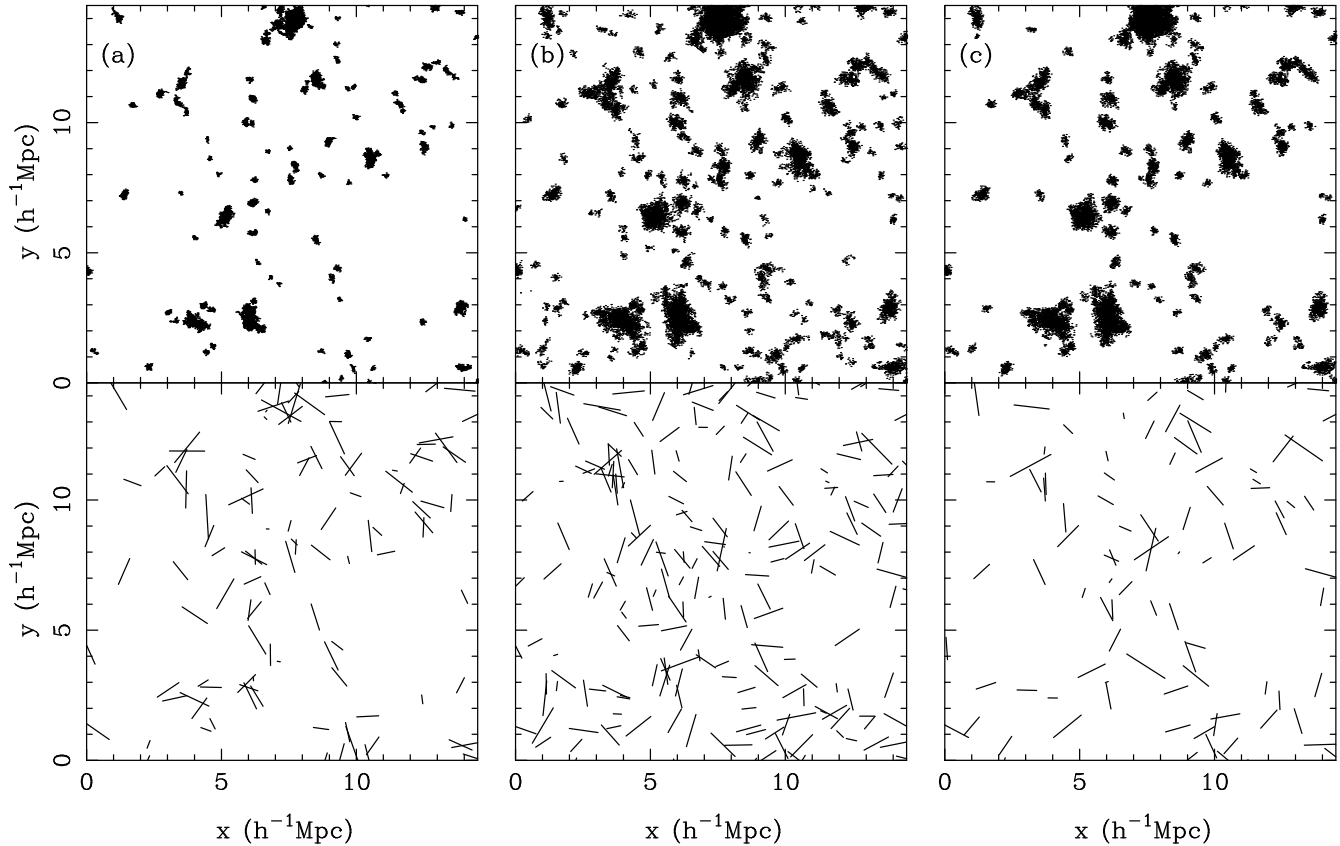

\centering
\PSbox{haloa.ps angle=-90 voffset=375 hoffset=-180 vscale=65 hscale=65}
{2.2in}{4.5in}
\PSbox{halob.ps angle=-90 voffset=375 hoffset=-180 vscale=65 hscale=65}
{2.2in}{4.5in}
\PSbox{haloc.ps angle=-90 voffset=375 hoffset=-180 vscale=65 hscale=65}
{2.2in}{4.5in}
\caption{
The (x,y) positions of particles which fall in halos (upper panels). We plot
particles found in a portion of the simulation box of (z) depth $40 \hmpc$.
Three different group finding techniques were used: (a) friends-of-friends
with a linking length of 0.1 times the mean interparticle separation
(FOF$_{0.1}$),
(b) friends-of-friends
with a linking length of 0.2 times the mean interparticle separation
(FOF$_{0.2}$),
and (c) the HOP groupfinder.
In the lower panels, we show the x and y components of the 
ellipticity of each halo (see text). The scale is such that an ellipticity of 1
corresponds to a bar of length $2 \hmpc$.
\label{halo}}
\end{figure*}

We use two different simulations, one run with a box size of $141.3 \hmpc$,
and another in a box of size $240 \hmpc$, both with $256^3$ particles.
The mass per particle in the former case is $1.4 \times 10^{10} \msun$,
and in the second it is 5 times larger. 
The Nbody code used was an adaptive particle-particle particle-mesh
code (Couchman, Thomas and Pearce), and the gravitational softening
length was $30 \hkpc$ (for the smaller box). 
For more details, the reader is referred to the Virgo papers,
including Kauffmann \etal (1999) for the smaller box simulation.
Unless stated otherwise, all our analysis will be carried
out using the higher resolution simulation, with the other being used as
 a check of the effects of lower resolution.
We use only the $z=1$ output in each case, as we will be interested in 
comparing to lensing observations where the peak of the galaxy distribution
 is  expected to lie close to this redshift.

\subsection{Halo finding}

Our analysis will be restricted to dark matter halos, and we
will make no attempt to identify luminous galaxies in the simulations.
In order to keep interpretation of our results simple, we will focus
mostly on halos picked from the particle distribution using one
easy to use groupfinder, the friends-of-friends (FOF) algorithm
(e.g., Davis \etal 1985).
A quick comparison will also be made with results from the 
publically available HOP groupfinding code (Eisenstein and Hut 1998).

In FOF, the one free parameter is the linking length as a fraction
of the mean interparticle separation, $b$. This in a crude way governs
the overdensity at the edge of the halo. We present some results
using either $b=0.1$, or $b=0.2$. The latter is often 
used so that particles will be chosen from regions roughly inside the
virial overdensity theshold of halos. With the former we will be
picking out the denser (by approximately a factor of $8$) central
parts of halos.

The HOP groupfinder works in a different way, using a smoothed
overdensity field calculated at the positions of the particles.
A group merging stage places subgroups together which fall inside
a certain density contrast. The HOP algorithm has 6 free parameters;
in this paper we use the fiducial values recommended by Eisenstein and Hut
(1998).

In Figure \ref{halo}, we plot the particles which ended up inside halos
for each of the three groupfinding choices (FOF$_{0.1}$, FOF$_{0.2}$,
and HOP). Only a small fraction of the box is shown, centred on an 
overdense region (the density inside the volume plotted, which measures
$15\times15\times40 \hmpccc$ is 1.6 times the cosmic mean). A lower
mass cut of 20 particles ($2.8 \times 10^{11} \msun$)
has been applied to the halo list, which results
in there being 20569 FOF$_{0.1}$ halos in the whole simulation
volume, 42024 FOF$_{0.2}$ halos, and 18023 HOP halos. 
Of course, the FOF$_{0.1}$ halos will tend to coincide with the 
central regions of the other halos, so that if we define the total
mass of a halo to be its virial mass, then this will be larger for the
FOF$_{0.1}$ objects than the number of particles which we plot in
Fig \ref{halo}. The FOF$_{0.1}$ halos are interesting because they 
will offer a good check of the robustness of our results, when we
ask below whether the ellipcities of the dense central regions are
 correlated in the same way as the outer parts of the halos.

\subsection{Halo ellipticities}

Although it is possible to measure three dimensional
ellipticities for halos,
 we will work here only with quantities which are projected
onto the plane of the sky, for ease of comparison with observational
data.

We estimate the ellipcity components of each halo by measuring the 
second moments of the projected mass distribution in a way
analogous to that used for the surface brightness distribution of galaxies
(Valdes \etal 1983, Miralda-Escud\'{e} 1991):

\begin{equation}
e_{1} = \frac{I_{xx}-I_{yy}}{I_{xx}+I_{yy}}, \qquad
e_{2} = \frac{2I_{xy}}{I_{xx}+I_{yy}}, 
\label{e1}
\end{equation}

where

\begin{equation}
I_{xx} = \frac{\sum_{i=1}^{N}(x_{i}-\overline{x})^{2}}{N}, \qquad
I_{yy} = \frac{\sum_{i=1}^{N}(y_{i}-\overline{y})^{2}}{N}, 
\end{equation} 

and

\begin{equation}
I_{xy}= \frac{\sum_{i=1}^{N}(x_{i}-\overline{x})(y_{i}-\overline{y})}{N}.
\end{equation}

Here, $\overline{x}$ and $\overline{y}$ are the coordinates of the
center of mass of the halo in question,
which contains $N$ particles of equal mass.
In the observational case, these moments are often calculated using the
surface brightness weighted by a function chosen to maximize the
S/N of the measurement (see e.g., VW). In our case, as we have mentioned 
above, we will probe the sensitivity of our measurements to variations
of this sort by comparing halos chosen with a large FOF linking length
to those using a small value which effectively give a non-zero weight
only to the dense central regions.

When working with pairwise statistics, such as the ellipticity correlation
function (see \S3), it will be advantageous to
redefine $e_{1}$ and $e_{2}$ for each pair of halos, 
with the $x$-axis defined to be line joining the halo centres and the $y$-axis
a line perpendicular to it. 

The quantity $(e_{1},e_{2})$ defines a pseudovector, of 
length $e=\sqrt{e_{1}^{2}+e_{2}^{2}}$. A positive $e_{1}$ component 
indicates a stretching along the $x-$axis, and a negative component
a stretching along the $y-$axis. The $e_{2}$ component is likewise a measure
of the stretching along axes at $45 \deg$ to the $x$ and $y$ axes. If we
have an ellipse of ellipticity $e$ ($e \equiv (1-q^{2})/(1+q^{2})$,
where $q$ is the ratio of minor and major axis lengths), then
\begin{equation}
e_{1}=e \cos(2 \beta), \qquad e_{2}=e \sin(2 \beta),
\end{equation}
where $\beta$ is the angle between the major axis and the $x-$axis.

In the bottom three panels of Fig. \ref{halo}, we plot the ellipticities
of the halos chosen by the three groupfinders. For each halo, we show
a bar with length proportional to $e$, oriented along the direction of the
major axis. It is difficult to pick out by eye any tendency for halos to be 
aligned. One worry which we might have had concerns the tendency of
close halos to be joined together by thin ``necks'' of particles and for
the groupfinder to count such pairs as one halo. If these halos 
are distributed along filaments, then we might expect an artificial 
alignment to arise. This does not seem to be an obvious problem here. For 
example, if we look at halos picked out by  the FOF$_{0.1}$
and FOF$_{0.2}$ groupfinders,
the ellipticities and directions appear to be fairly similar even though
different parts of the halo are being used to calculate them. We will check the
statistical tendencies for alignments in the next section. The HOP halos
do in several cases appear to be made from several small halos joined together.
Presumably this could be alleviated by tuning the free
parameters which govern the method, but we have not attempted to do this.

\begin{figure}
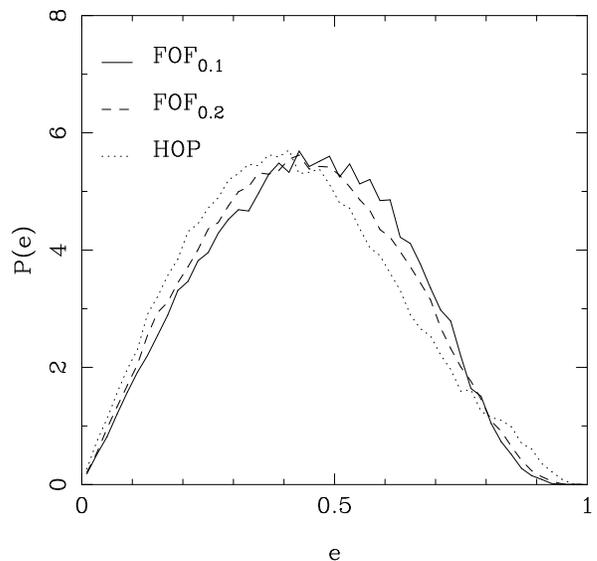

\centering
\PSbox{ehist.ps angle=-90 voffset=235 hoffset=-35 vscale=42 hscale=42}
{3.5in}{3.0in} 
\caption
{
The probability distribution function of halo ellipticities, for the 3
different group finding techniques.
\label{ehist}
}
\end{figure}
 
In Fig. \ref{ehist}, we show the probability distribution of $e$ values
for the halos (again with $>20$ member particles) picked out by the three
groupfinders. They are all fairly similar, with a mean $e$ around $0.4$, 
slightly higher than the $0.3$ typical of real galaxies (e.g.,
Wittman \etal 2000).

\section{Ellipticity correlations in three dimensions}

Observationally, ellipticity correlations are measured as a function
of angular separation, on the plane of the sky (at least for weak 
lensing surveys). With distance information (for example, using redshifts),
 it would be possible 
to measure them as a function of separation in three dimensions, and this
is what it is most natural to do using our Nbody simulations. In this section,
we will do this, and later convert the measurements to angular correlations
which can be compared to weak lensing survey results. This conversion
will be done in two different ways. The first is an analytical projection
of our three dimensional results using Limber's equation (Limber, 1959).
The second is a direct measurement from simulated surveys made by
 projecting the halo distributions in the box, and applying a radial 
selection function.
In the present section, although we will be dealing with halo separations
in three dimensions, it is worth bearing in mind that we  restrict
ourselves to quantities which can be measured directly observationally
(albeit with redshifts), so that the ellipticities we will be 
correlating  are projected ellipticities (defined in Equation \ref{e1}).

Two ellipticity correlations can be defined (following
 Miralda-Escud\'{e}  1991) as
\begin{equation}
c_{11}(r)=\langle e_{1}({\bf x}) e_{1}({\bf x}+ {\bf r}) \rangle,
\label{c11}
\end{equation} 
and
\begin{equation}
c_{22}(r)=\langle e_{2}({\bf x}) e_{2}({\bf x}+ {\bf r}) \rangle,
\label{c22}
\end{equation} 
together with a cross-correlation,
\begin{equation}
c_{12}(r)=\langle e_{1}({\bf x}) e_{2}({\bf x}+ {\bf r}) \rangle,
\label{c12}
\end{equation} 
where ${\bf r}$ is the three dimensional vector of length
$r$ joining each pair of halos.
Here we have defined the ellipticity components, $e_{1}$ and $e_{2}$
with respect to axes which are the projection into the $x-y$
plane of the line joining each pair of halos and a line orthogonal
to it which lies also in the x-y plane.

 In general, these
functions may  be anisotropic, and depend on the separation
between pairs both across and along the line of sight. We will plot their
full dependance on these quantities later. In our tests of different halo
finders and simulation resolution, we will work with $c_{11}(r)$ only (although
we have checked that our conclusions are valid for the other functions also).
 This is because $c_{11}$ has no strong angular dependence, so that plotting
it as a solely a function of $r$ does not hide any crucial aspects.

\begin{figure}[t]
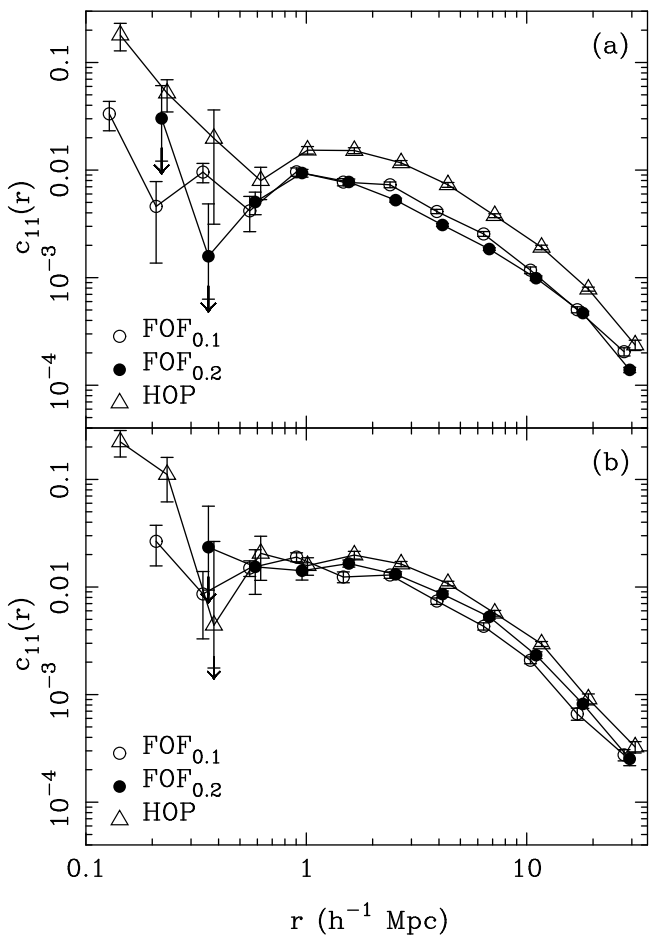

\centering
\PSbox{xie.ps angle=-90 voffset=420 hoffset=-150 vscale=70 hscale=70}
{3.5in}{5.5in} 
\caption
{
The ellipticity correlation function $c_{11}(r)$ (Equation \ref{c11}) 
as a function of pair separation
between halos in real space. In panel (a) we show results for all
halos, for three different groupfinders (as in Fig.~\ref{halo}).
 In (b) we show results for halos
of the same space density (we apply a minimum mass cut in each case and use
the most massive 10000 halos in the simulation- see text).
For clarity, the points have been displaced slightly in the $x$ direction.
\label{xie}
}
\end{figure}

We show $c_{11}(r)$ in Fig. \ref{xie}, for the three different groupfinders.
In the top panel, we give results for all halos with $> 20$ member 
particles. The error bars are the equivalent of Poisson errors, calculated
by randomly rotating the halo major axes and the calculating $c_{11}(r)$ for 
this randomized distribution (see e.g.,  VW).
 The dispersion between results for 
several such randomizations (we use 50) gives the error bar. In order to 
make best use of all the information contained in the simulation,
we have averaged results from the three orthogonal projections of the halo 
particles.

The first thing to notice is that there is a good detection of a signal, even 
for the largest bin that we plot. The widest pairs we use have a separation
of 0.25 times the box side-length, or $35 \hmpc$. This is not 
necessarily that surprising, given that large scale correlations in the 
matter distribution exist on such scales and beyond.

The FOF$_{0.1}$ and FOF$_{0.2}$ halos give very similar results. As mentioned
earlier, this is a good test of the robustness of the correlation
signal and a good argument that there are not systematic problems. It is 
telling us that the inner and outer parts of halos are responding in the
same way, a stability which points to the correlation being a real effect.
The HOP results are larger by a factor of $\sim 2$, and on small scales
seem to deviate even more. We have already seen from Fig. \ref{halo} that
with the free parameters set in the way we have chosen,
 this groupfinder seems to produce groups
which are clumps of subgroups which the FOF algorithm keeps separate.
This is probably an indication that for a more reasonable
result, we would need to tune some of the six free parameters
that govern HOP.

Some of the differences between $c_{11}(r)$ results might be due to the fact
that the same absolute mass cut (20 particles) for all three groupfinders
will result in different halos being chosen. For example, the FOF$_{0.1}$
halos will be much rarer than 
FOF$_{0.2}$. A better way to compare results is to
set the space density of halos to be equal. We have done this the bottom
panel of Fig.\ref{xie}, where we show $c_{11}(r)$ for the most massive
10000 halos in the simulation volume. The lower mass limits for the different
groupfinders in this case are 44, 90 and 85 particles
($6.2\times 10^{11} \msun$, 
$1.3 \times 10^{12} \msun$ and $1.2 \times 10^{12} \msun$),
 for FOF$_{0.1}$,
FOF$_{0.2}$ and HOP, respectively. The results are indeed more similar now,
except for the HOP results at the smallest separations.
We have also tried applying an upper mass cut, keeping only halos
containing less than a certain number of particles. We find similar results
(not plotted) to those for all halos.

For the rest of the paper, we have decided to use groups
chosen by the FOF$_{0.1}$ groupfinder. This is because we are most interested
in the inner parts of halos, which is where galaxies will tend to lie.
Also, when halos are chosen to have the same space density (Fig. 
\ref{xie}b), these halos have a slightly lower correlation than the 
others, so that we are being conservative.

\begin{figure}
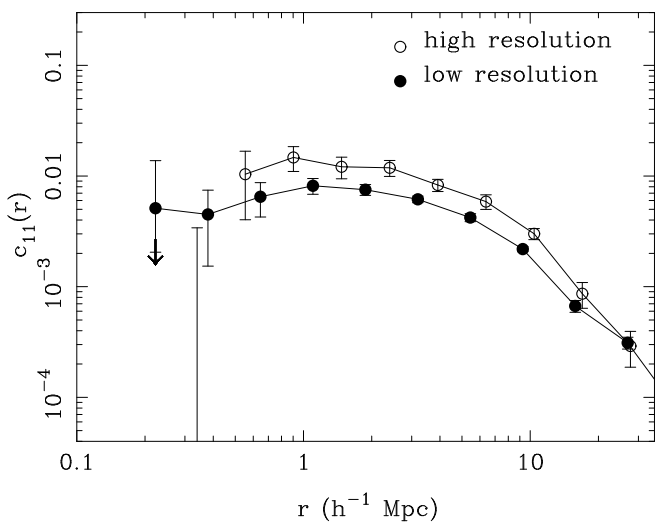

\centering
\PSbox{xieres.ps angle=-90 voffset=270 hoffset=-55 vscale=48 hscale=48}
{3.5in}{3.0in} 
\caption
{
The ellipticity correlation function $c_{11}(r)$ (Equation \ref{c11}) 
as a function of pair separation
between halos in real space. We show results for halos of the same
mass (which also have the same space density) taken from simulations with
two different spatial and mass resolutions (see text). The lower resolution
simulation is larger in volume by a factor $\sim5$, and has a mass resolution
$\sim 5$ times lower. For this figure, and the rest of the paper,
we use FOF$_{0.1}$ halos.
\label{xieres}
}
\end{figure}

One can ask about the effect of simulation resolution on our results, 
whether the simulations have converged and if not, whether lower resolutions
yield higher or lower correlations. In Figure \ref{xieres}, we compare
halos taken from the larger, low resolution
 simulation (240 $\hmpc$ boxsize) to our 
previous results. We again use the FOF$_{0.1}$ groupfinder, and in order
to compare halos with the same mass, we use halos containing
$>20$ particles in the low resolution case and $>100$ in the high resolution 
case (these halos do in fact have the same space density, indicating that 
the simulations have converged as far as the mass function is concerned).

We can see that although the ellipticity correlations have not converged,
the higher resolution simulation is more correlated. This may be because 
structure has been allowed to form on smaller scales, whereas if the
the structure is unresolved, some of the tidal field which causes 
correlations to arise is missing. In any case, we can argue that 
the fact that higher resolution increases the correlated signal means that 
our results are likely to be conservative. Of course this situation is far
from ideal, and one would really like to see if even higher resolution
results do in fact converge. Such simulations are beyond the 
scope of this paper, however. This is particularily true because it
is important not to sacrifice a cosmologically interesting volume
for the sake of higher resolution, since long range tidal forces may play a
significant role in setting up the correlations.

\begin{figure}[t]
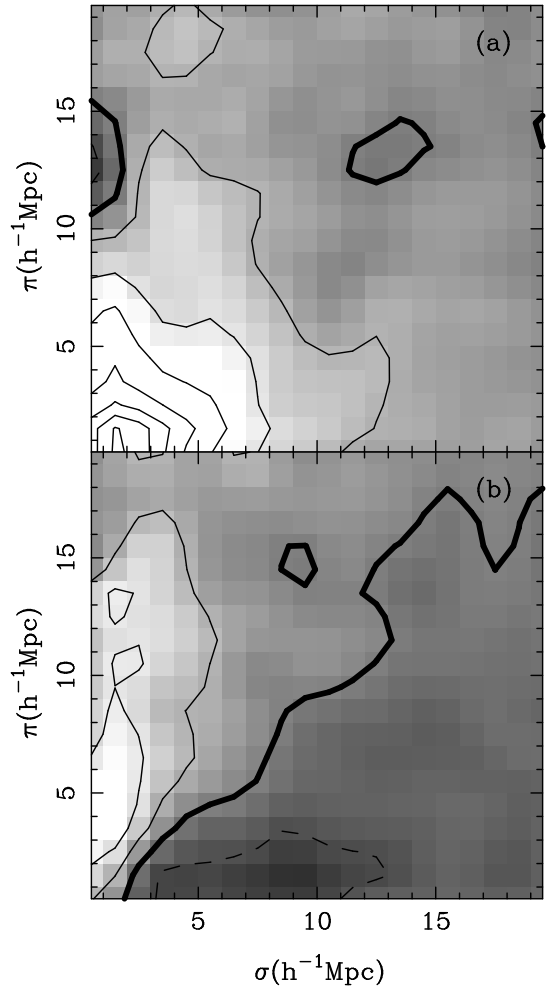

\centering
\PSbox{sigmapixitr.ps angle=-90 voffset=430 hoffset=-170 vscale=75 hscale=75}
{3.5in}{5.5in} 
\caption{
The ellipticity correlation functions, (a) $c_{11}(\sigma,\pi)$, 
and (b) $c_{22}(\sigma,\pi)$.  Here
$\sigma$ is the distance between pairs of halos perpendicular to the line of
sight, and $\pi$ is the distance parallel to the line of sight.
The values of  $c_{11}$ and $c_{22}$ were originally binned into $20\times20$
bins in $\sigma \times \pi$. For clarity, we have plotted these results
smoothed with a Gaussian window with dispersion $\sigma_w=1$ bin.
The contours represent increments of 0.001 in $c_{11}$ and $c_{22}$, with the 
thickest contour representing $0$ and dashed contours negative 
values.
\label{sigmapixitr}
}
\end{figure}

In Fig. \ref{sigmapixitr}, we show $c_{11}$ and $c_{22}$ as a function 
of halo pair separation across the line of sight ($\sigma$) and 
along the line of sight ($\pi$). This presentation is similar to that 
used in plotting redshift distortions of the matter correlation 
function (e.g., Kaiser 1987). Only $c_{22}(\sigma,\pi)$ appears to be
severely anisotropic. The plot seems to be showing us that projected halos tilt
in opposite directions (-ve correlation) when they are side-by-side,
but in the same direction when they lie one behind the other. The
$c_{11}(\sigma,\pi)$ function is nearly isotropic, and
positive, so that pairs of halos have a tendency to stretch 
along the projection of the line  joining them whatever their spatial
orientation. 

 If $c_{22}$ is measured averaged in radial bins,
$r$ (where $r=\sqrt{\sigma^{2}+\pi^{2}}$) then the signal from the 
bottom right  partly cancels that in the top left and the result is 
a very small correlation function.
 On the other hand, a projection of $c_{22}$ on the plane
of the sky (which we will do later when calculating the angular
correlations) is equivalent to a projection along the $\pi$ axis. Because
$\pi$ cannot be negative (it is defined as the modulus of the pair
separation along the line of sight), then the contribution of
the positive $c_{22}$ signal at large $\pi$ will result in 
a significant angular signal.

We do not plot the cross-correlation, $c_{12}(\sigma,\pi)$ here, as 
we find it to be consistent with zero everywhere. We also note that 
in the observational case, plots like these (and the previous plots which 
were radially averaged) will be affected by redshift distortions
of the pair separations.

\section{Simulated surveys}

One way of making predictions for the ellipticity correlations that can be
compared directly to current observations is by using simulated 
surveys with a similar geometry to real surveys. In this section
we  describe our creation of such datasets from the Nbody
halo catalogs. In \S5 we measure angular statistics
from the simulated surveys and compare to analytic projection
of the three dimensional statistics of \S3.

The redshift distribution of faint background galaxies that are used in
current weak
lensing studies is not known accurately. Most of the observational
groups assume that the peak of the redshift histogram 
lies around $z=1$, so we will use this value in setting up our simulated
surveys. We note that depending on the actual redshift of the galaxies, 
the relative contribution and angular scale of the lensing correlations
and any intrinsic correlation will vary, something which can undoubtedly
used to tell them apart. In our mock surveys, we will also assume that 
the galaxy population
 (we refer to halos as ``galaxies'' when describing the
mock surveys) stays constant in comoving coordinates, so that we only
make use of the $z=1$ simulation output. 
We leave the study of the effects of redshift evolution to future work.
Also, this approach will allow us to compare directly with the analytic
projection of the statistics.

Following what has now become a standard technique (e.g., Jain, Seljak
\& White 2000,
Croft \etal 2000) we stack simulation boxes one behind the other, from $z=0$
to $z=2$. Each simulation box is subjected to a random recentering,
and possibility of reflection about one of the three centres. We place
 each
with one of the three axes (randomly chosen) pointing along the line of sight.
This procedure is adopted to avoid periodic repetition of structures.
When projecting the box contents, we work in comoving coordinates,
and set the angular size of the simulated catalog to be one box width
at $z=1$, where our $z$ histogram will peak. At $z=1$, an angular scale
of $1\deg.$ therefore corresponds to $40 \hmpc$. We make use of the small
angles involved by projecting onto a flat plane, rather than a curved sky.
This facilitates things because the halo ellipticities have been defined
in the same fashion.

The $N(z)$ form we choose for our mock surveys is a Gaussian 
distribution 
 $N(z)=e^{(z-1)^{2}/(2\sigma_{z}^{2})}$ (for simplicity's sake),
where we have centered on $z=1$. We apply the
selection function $\psi(r)$ which leads to such
an $N(z)$ for a uniform distribution of galaxies, where
$\psi(r)$ is the relative probability of including randomly chosen galaxy
at a comoving distance $r$ (see e.g., Peebles 1993).
We try 3 values for $\sigma_{z}$, $0.4,0.2 $ and $0.1$. 
The widest of these is designed to mimic roughly the present observational
catalogs which use no colour selection. The narrower redshift
distributions could arise if some sort of photometric
redshifts are used (narrower distributions still are possible,
 see e.g., Hogg \etal 1998).

It is necessary to project 26 box lengths to reach $z=2$. For galaxies
past the peak in the $N(z)$, the mock survey is wider than the simulation
box. At this point we make use of the periodic boundaries and make each layer 
of boxes periodic across the line of sight. This was also done by 
e.g., Seljak, Burwell \& Pen 2000) and should not cause any problems. In 
any case, we will compare with 
the analytic projection of the statistics to make sure.

\begin{figure}
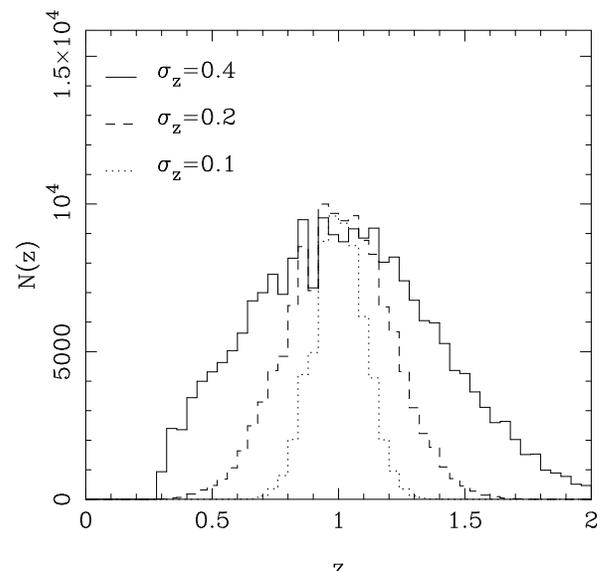

\centering
\PSbox{nofz.ps angle=-90 voffset=235 hoffset=-35 vscale=42 hscale=42}
{3.5in}{3.0in} 
\caption
{
The actual redshift distribution of galaxies in simulated catalogs
generated with different values of $\sigma_{z}$. We plot the number
of galaxies in bins of width $\Delta_{z}=0.04$
\label{nofz}
}
\end{figure}

We use only the FOF$_{0.1}$ halos (with $>20$ particles) to make our simulated
surveys. This results in a rather small number of
galaxies per square degree, $\sim 18000$ 
(for $\sigma_{z}=0.4$), compared to
e.g., $\sim 10^{5}$ for VW .
The $N(z)$  for three surveys made using different value of $\sigma_{z}$
is plotted in Fig. \ref{nofz}. 
We set the selection function to zero for
$z<0.3$ and $z>2$.

 \begin{figure}[t]
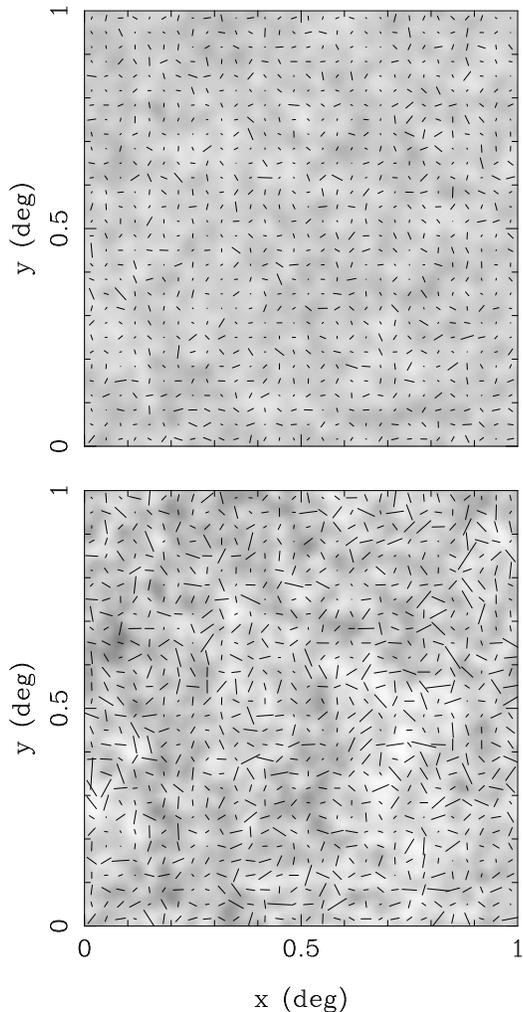

\centering
\PSbox{simskya.ps angle=-90 voffset=230 hoffset=-55 vscale=45 hscale=45}
{3.5in}{3.0in} 
\PSbox{simskyb.ps angle=-90 voffset=230 hoffset=-55 vscale=45 hscale=45}
{3.5in}{2.5in} 
\caption
{
A $1\deg \times 1 \deg $ patch of the simulated
sky. 1 $\deg$ corresponds to a comoving size
of $40 \hmpc$ at $z=1$, where the 
galaxy distribution peaks. The upper panel is for a survey where the galaxies
have a wide redshift distribution ($\sigma_{z}=0.4$), and the lower panel
has $\sigma_{z}=0.1$. We have show the ellipticities averaged
 in square bins of width
2 arcmin, as bars, with a length
of $0.15 \deg$ corresponding to an ellipticity of 1. We also show the
projected halo density field smoothed with a Gaussian filter of FWHM 2 arcmin,
with a linear grayscale. 
\label{simsky}
}
\end{figure}

In Fig.\ref{simsky}, we show the $x$ and $y$ components of the ellipticities
on the plane of the sky, where we have binned the individual
$e_{1}$ and $e_{2}$ values in square cells of side-length 
2 arcmins. We plot as 
a line, the average ellipticity in each cell in a manner analogous to the
individual halo ellipticities plotted in Fig \ref{halo}. We also show, as a
grayscale, the angular area density of halos smoothed with a Gaussian 
filter of FWHM 2 arcmins. Fig. \ref{simsky}a represents a survey
with a wider $z$ distribution ($\sigma_{z}=0.4$) than Fig. \ref{simsky}b.
The obvious difference between the two is that the wider distribution
has washed out both the density inhomogeneities and the absolute value
of the residual ellipticities. To the eye, there does not seem to be any
obvious correlation between the ellipcitity vectors, although
as we will see in the next section, there are detectable statistical
correlations and a non-zero shear variance. We defer any exploration
of the relation between the projected halo density and ellipticities
to future work.

\section{Angular statistics}

\subsection{Angular ellipticity correlations}

The ellipticity correlations measured in three dimensions
in \S3 can be projected for direct comparison with results from 
angular galaxy surveys. This can be done simply using a
modification of Limber's
equation (Limber 1953, Peebles 1993), so that the angular
correlation function of $e_{1}$ components of pairs is  
\begin{equation}
C_{11}(\theta)=\frac{\int (r_{1}r_{2})^{2} dr_{1}dr_{2}\psi_{1}\psi_{2}
(1+\xi(r_{12})c_{11}(\sigma,\pi)}
{[\int r^{2} dr \psi]^{2} + \int(r_{1}r_{2})^{2}dr_{1}dr_{2}\psi_{1}\psi_{2}
\xi(r_{12})},
\label{limber}
\end{equation}
where $r_{12}$ comoving is the separation between points at distance
$r_{1}$ and $r_{2}$ from the observer. 
$\sigma$ and $\pi$ are the separations across and along the
line of sight, which we calculate in the small angle limit,
so that 
\begin{equation}
\sigma=(\frac{r_{1}+r_{2}}{2}) \theta, \qquad \pi=\mid(r_{1}-r_{2})
(1-\theta^{2}/2)\mid,
\end{equation}
and $r_{12}=\sqrt{\sigma^{2}+\pi^{2}}$. The selection functions at
$r_{1}$ and $r_{2}$ are given by $\psi_{1}$ and $\psi_{2}$, and $\xi(r)$
is the halo autocorrelation function. The relations for the
$e_{2}$ component correlation, $C_{22}(\theta)$ and the cross-correlation,
$C_{12}(\theta)$ are analogous.

\begin{figure}[t]
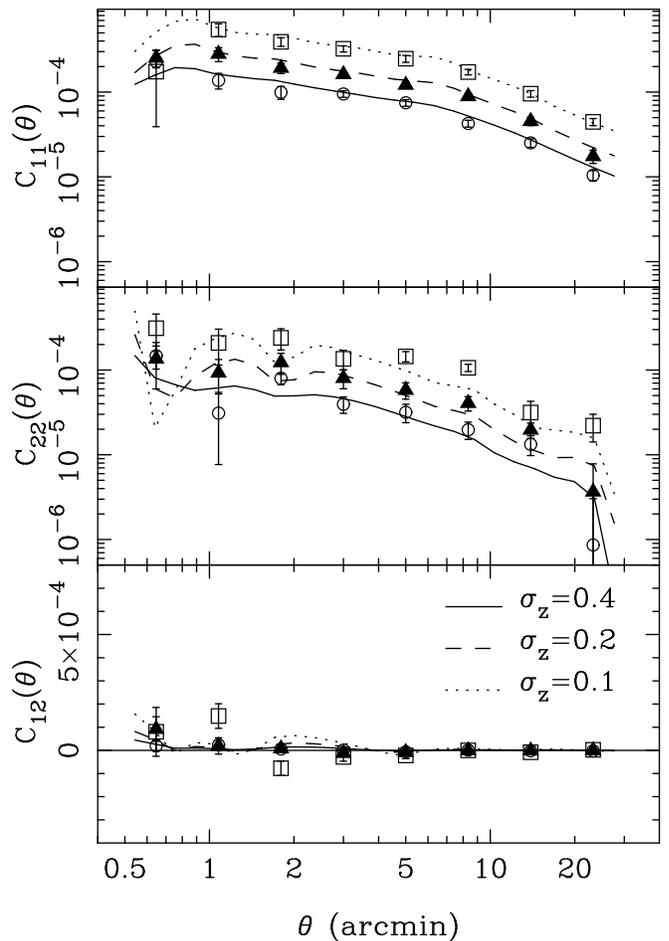

\centering
\PSbox{simcatvslimber.ps
 angle=-90 voffset=405 hoffset=-150 vscale=70 hscale=70}
{3.5in}{5.25in} 
\caption
{
Comparison of angular ellipticity correlation functions 
computed using Equation \ref{limber} (lines) and directly from simulated 
catalogs (points). We show correlations of the 
$e_{1}$ components ($C_{11}$) the $e_{2}$ components ($C_{22}$) and
the cross correlation of components ($C_{12}$),
 as a function of pair separation
between halos in angular coordinates.
Note that in the bottom panel, the y-axis is on a linear scale.
\label{simcatvslimber}
}
\end{figure}

Using this formalism, we calculate 
$C_{11}(\theta)$, $C_{22}(\theta)$ and $C_{12}(\theta)$
for the three different selection functions also used in our simulated
surveys, which yield a Gaussian $N(z)$. The results are
plotted in Fig.\ref{simcatvslimber}. We can see that for the first two of these
a distinct correlation signal results. The behaviour as
a function of redshift width of the galaxy distribution is as we might predict,
with a wider distribution yielding less angular correlation. Also very 
interesting is the fact that $C_{11}(\theta)$ is everywhere positive, whereas
$C_{22}(\theta)$ dips below zero on the largest scales. 
The $C_{12}$ function is much smaller, oscillating about zero, due to noise.
These functional
forms are very similar to the theoretical expectation for the
weak lensing signal (see e.g., Miralda-Escud\'{e} 1991, Kaiser 1992). A simple 
explanation  of this similarity is that the shear field which 
is responsible for the weak lensing correlations is mathematically
similar to the shear field responsible for generating galaxy intrinsic
shapes and spins from the surrounding large-scale structure, something
which can be used to make fully analytic predictions of the effect, in
the context of linear theory (Kamionkowski, private communication).
We will return briefly to this in \S7.

We have put points on Fig \ref{simcatvslimber}
which come from directly estimating the correlation 
functions from our simulated surveys of \S4. We use as estimators
the angular analogs of equations \ref{c11}-\ref{c12}.
The error bars have been computed from
the error on the mean measured from 10 simulated surveys set up with different
random seeds. As we use the same underlying Nbody simulation for all, this
will result in an underestimate of the cosmic variance. This is adequate for
our purposes, because we are comparing to the Limber's equation 
projection of the three dimensional
statistics which is not likely to be exact anyway,
 as it involves numerical intergration
of a noisy function. This is particularily true of $C_{22}(\sigma,\pi)$.
That said, the agreement between the two sets of  angular statistics is 
good, so that we can be fairly confident that there are no serious errors
in the conversion of three dimensional statistics to angular ones. 

\subsection{Shear variance}

The variance of the galaxy ellipticities binned into cells on the plane of
the sky is a simple and interesting statistic to calculate. An
estimator for the shear in a cell centred on angular
 position ${\bf \phi_{i}}$ is
(following VW, although note that we assume 
that the shear, $\gamma=\langle e \rangle /2$,
 in the limit of intrinsically round galaxies, so that 
our definition is slightly different) is
\begin{equation}
E[\gamma^{2}({\bf \phi_{i}})] =
[\frac{1}{2N} \sum_{j=1}^{N} e_{1,j}]^{2} +
[\frac{1}{2N} \sum_{j=1}^{N} e_{2,j}]^{2},
\label{estimator}
\end{equation}
where there are $N$ galaxies in the cell and $e_{1,j}$ and $e_{2,j}$
are the ellipticity components of the $j$th galaxy, which have been defined
with respect to some fixed axes. The shear variance is then 
 $\langle E[\gamma^{2}({\bf \phi_{i}})] \rangle= (\sigma^{2}_{e}/4N) +
\langle \gamma^{2} \rangle$. Here $(\sigma^{2}_{e}/4N)$ is an ellipticity
shot noise term, arising from the random intrinsic
ellipticities of the finite number of galaxies involved,
 and $\langle \gamma^{2} \rangle$ is the quantity we
are interested in. 

\begin{figure}[t]
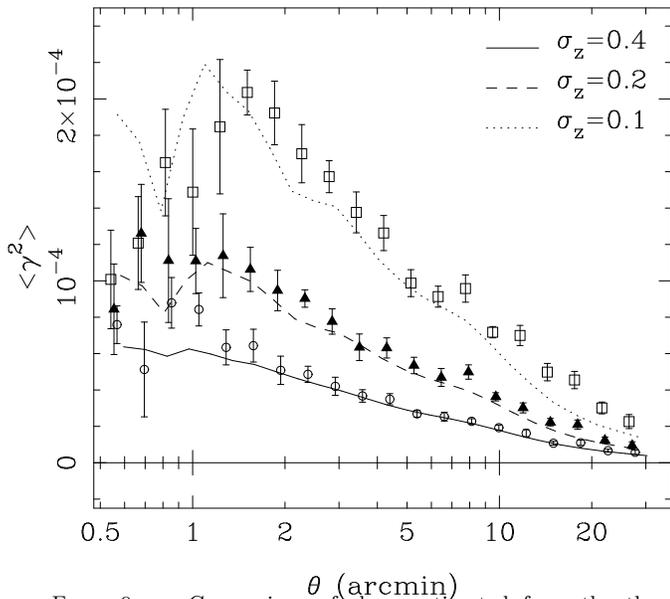

\centering
\PSbox{limbervsimcatshear.ps
 angle=-90 voffset=255 hoffset=-55 vscale=48 hscale=48}
{3.5in}{3.0in} 
\caption
{
Comparison of shear estimated from the three-dimensional
ellipticity correlation (Equations \ref{limber} and \ref{shearint})
 and results measured
directly from from simulated 
angular surveys (see \S5.2).
We show results for three different redshift widths of the $N(z)$
distribution (see Fig.\ref{nofz}).
\label{limbervsimcatshear}
}
\end{figure}

In terms of the angular ellipticity correlation functions,
 $\langle \gamma^{2} \rangle$ is given by a double integral over the 
cell being used (as in the relation between $\xi$ and counts-in-cells 
variance, e.g., Peebles 1993), so that 
\begin{equation}
\shear=\frac{1}{4A^{2}}\int \int C_{11}(\theta) + C_{22}(\theta) d^{2}A,
\label{shearint}
\end{equation}
where $A$ is the cell area.
We note that this integral will tend to be dominated by the largest 
$\theta$ scale, so that sensitivity to systematic errors on small 
scales will be relatively low.

We carry out this integration using our angular ellipticity correlations
from Fig. \ref{simcatvslimber}. For easy comparison with our simulated survey
results (see below), we use square cells of side length $L$. In Figure
\ref{limbervsimcatshear}, we show the $\shear$ curves that result from the 
integral, as a function of angular scale $\theta=L/\sqrt{\pi}$.
 This $x$-axis scaling was chosen because some obseravtional groups 
(e.g., VW) calculate $\shear$ in round cells of radius $\theta$.

From the plot, we can see that $\shear$ rises as we move from large to small 
scales. The rms shear, $\shear^{1/2}$ at $\theta=2$ arcmins
is $\sim 0.6 \%$ for a wide redshift distribution ($\sigma_{z}=0.4$).
This is significantly below the theoretical prediction 
for the rms shear due to weak lensing in different
cosmological  models, which can vary
from $\sim1\%$ to $>3 \%$ (see e.g., VW). Because the
contributions to this quantity add in quadrature,
the intrinisic shear is likely only to matter  for
low amplitude models. Of course, as we have only
simulated one (relatively high amplitude) model, it could be that the
intrinisic shear for low amplitude cosmologies is also lower.
For a narrower redshift distribution with $\sigma_{z}=0.1$,
our value of $\shear^{1/2}$  doubles.
On large scales we also see that 
there is still signal out to $\theta=30$ arcmins.

For comparison, we again show points representing the value of
$\shear$ measured from the simulated surveys. We have used the estimator of
equation \ref{estimator},
 with square cells of side length $L$ ($=\sqrt{\pi}\theta$).
To remove the shot noise term, we use the procedure advocated 
by VW. This involves randomly rotating the major axes of all the projected
halos and calculating $\shear$, which will be a noisy version of
the shot noise term. We carry out 1000 such randomizations and use the
average $\shear$ measured from them as our $\sigma^{2}_{e}/4N$, which
we subtract. The results are shown in Fig. \ref{limbervsimcatshear},
with error bars which are again the error on the mean from 10 simulated
surveys. As with the ellipticity correlation functions, the agreement
is good.

\section{Comparison with observations}

As mentioned in 
\S1, there are now several different groups 
 (VW;
 Bacon, Refregier \& Ellis 2000;
Wittman \etal 2000; KWL)
producing results
from large observational campaigns to measure weak lensing by large-scale
structure. 
These different surveys (at the time the published results were
submitted) involved from 0.5 to 1.8 square degrees of imaging data,
and produced measurements of ellipticity correlations and shear
on scales from $\sim 0.6$ to $\sim 30$ arcmins.
In this section, we will briefly compare these current
measurements with the predictions of \S5.

\begin{figure}[t]
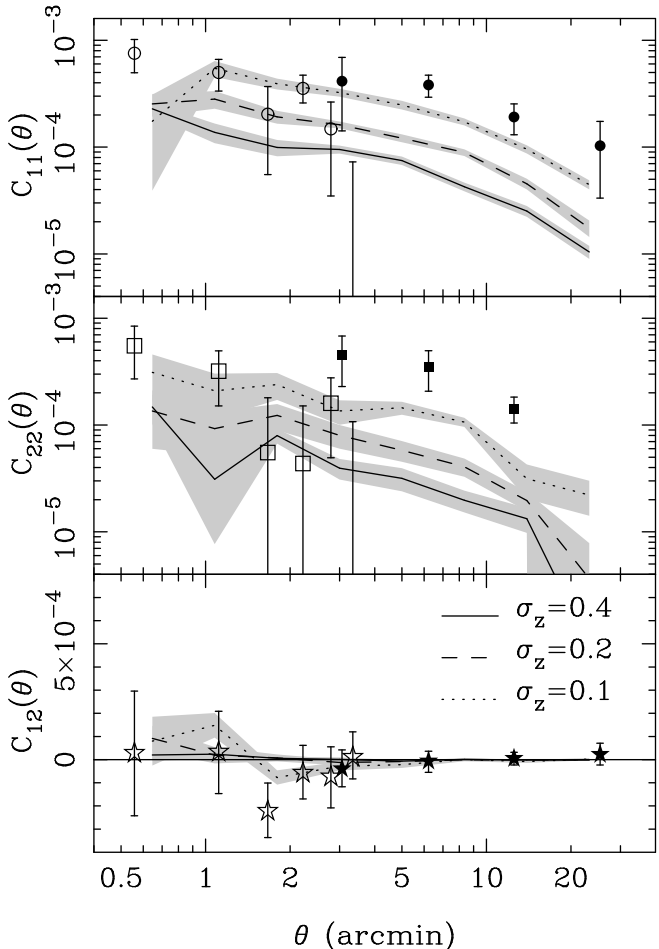

\centering
\PSbox{thetatrobs.ps angle=-90 voffset=405 hoffset=-150 vscale=70 hscale=70}
{3.5in}{5.0in} 
\caption
{
The angular ellipticity correlation functions
measured from galaxy survey data by VW (open symbols) and
Wittman \etal (2000) (filled symbols).
We also show the measurements made in \S5 from our simulated surveys,
for three different values of the redshift width ($\sigma_{z}$). The 
value of   $\sigma_{z}$ which is likely to be closest to that in the  
observed galaxy distributions is $\sigma_{z}=0.4$, which gives the lowest 
curves. We show the error bars
taken from Fig. \ref{simcatvslimber} as shaded regions. 
\label{thetatrobs}
}
\end{figure}

Two of these groups (VW, and Wittman \etal 2000) and have published 
 ellipticity correlation functions,
using slightly different notation from ours. What we call $C_{11}(\theta)$,
$C_{22}(\theta)$, are called 
$\langle e_{t}(0) e_{t}(\theta) \rangle$,
$\langle e_{r}(0) e_{r}(\theta) \rangle$ by VW, and
$\langle e_{1}e_{1} \rangle$, $\langle e_{2}e_{2} \rangle$ 
by Wittman \etal (2000).
The cross correlation $C_{12}(\theta)$ is defined analogously.

We show these observational results in Fig. \ref{thetatrobs}, together
with those from the simulated surveys (from Fig. \ref{simcatvslimber}).
The measurements from the two groups are mostly at different scales,
but agree within the errors in the small degree of overlap
around $\theta=3.5$ arcmins. The functional form of our intrinsic 
correlation follows roughly the observational results. The amplitude
is however much lower, by a factor of roughly between 5 and 10 for most of the 
points, for  the most relevant redshift distribution (a wide one, with 
$\sigma_{z}=0.4$). The Wittman $\etal$ points appear to be even higher
 than this on large scales for $C_{22}$. We note that in the
Wittmann $\etal$ paper, these points are shown to 
also be rather higher compared to theoretical weak lensing predictions
for $\Lambda$CDM than the $C_{11}$ points.
 For a narrower redshift distribution, which may be
achieved in the future with photometric redshifts, the intrinsic
correlations are closer to the current observational data.
The cross correlation, $C_{12}$, while noisy, is small, and consistent with 
zero in all cases.

\begin{figure}
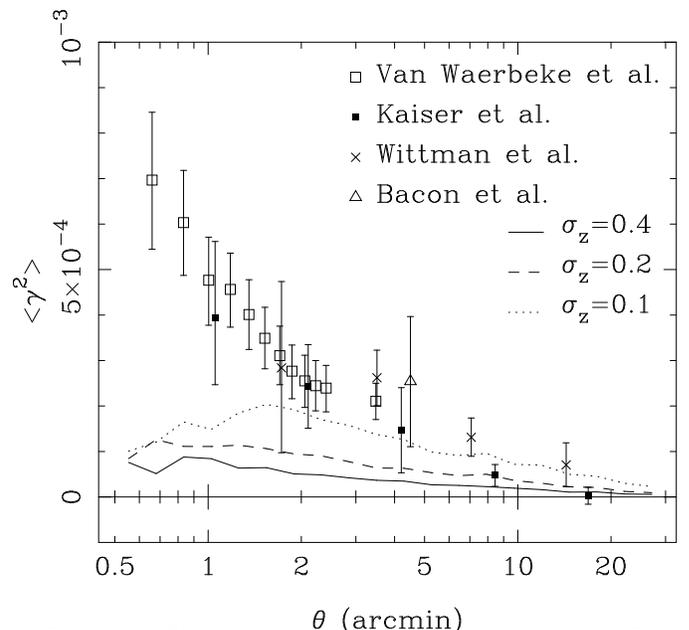

\centering
\PSbox{shearobs.ps angle=-90 voffset=255 hoffset=-55 vscale=48 hscale=48}
{3.5in}{3.0in} 
\caption
{
Observations of the shear variance from four different 
groups (points with error bars), together with results from our 
simulated surveys (lines). 
We show the latter for three different values of the redshift 
width ($\sigma_{z}$). The 
value of   $\sigma_{z}$ which is likely to be closest to that in the  
observed galaxy distributions is $\sigma_{z}=0.4$, which gives the lowest 
curves. 
The error bars on 
the simulation points can be obtained from Fig. \ref{limbervsimcatshear}.
\label{shearobs}
}
\end{figure}
 
If we move to the shear variance, $\shear$, we can plot results
from all four groups. We have the plotted these in the same format as
Figure 2 of KWL, as Fig.\ref{shearobs}.
As Wittman \etal did not specifically
publish $\shear$ values, we have converted their $C_{11}$ measurements using
the formulae in Kaiser (1992) (as was done in KWL,
 to which the reader is referred for details),
 assuming a spectral index of mass fluctuations of $n=-1$.
One difference between this plot and that in KWL
is that we plot the results as a function of $\theta$, the radius
of a top-hat sphere. To do this, we have converted the 
square box side-length $L$ from KWL to the radius of a circle
with the same area ($\theta=L/\sqrt{\pi}$; this is the
same as was done in plotting Fig. \ref{simcatvslimber}) .  
The groups all have results which appear to be consistent within their
errors, which is impressive and reassuring.

Turning to the simulation results,
 we concentrate first on the wide redshift distribution ($\sigma_{z}=0.4$),
which is most relevant to the current observations.
The intrinsic $\shear$ we predict is a small fraction of the measured value
on all scales, except possibly for the 
largest scale point from KWL, which, as those authors point out is
an interesting null detection. The intrinisic $\shear$
is also at a smaller level than the $1\sigma$ errors for the observational
measurements. It is therefore not really a factor when interpreting the
current results, but may become a bit more of an issue when the surveyed
areas increase and the statistical errors become smaller.
If we now turn to the other two curves, we can see that for the narrow
redshift distributions, the intrinsic $\shear$ begins to approach the
current measurements, at least on large scales. On small scales,
although the errors are large (Fig. \ref{limbervsimcatshear}),
there appears to be relatively less signal.

\section{Summary and Discussion}
We have used Nbody simulations to make predictions for
intrinsic correlations of galaxy ellipticities, under the assumption
that galaxy shapes follow the shapes of their dark matter halos. 
Measurements of the ellipticity correlation functions in three-dimensions
give a distinctive signal, which we measure with
relatively small uncertainties 
on scales from $\sim 0.5-30 \hmpc$. These correlations
vary by less than a factor of $\sim 2$ for different 
halo finding techniques and different simulation resolutions.
We project these three dimensional correlations into angular statistics,
including the shear variance. We have done this
 both analytically, using a modified Limber's
equation, and by making direct measurements from 
simulated surveys constructed by projecting the simulation
boxes. We find that the amplitude of the angular statistics 
depends strongly on the redshift width of the galaxy distribution. With
widths appropriate to present day surveys, we find that the intrinsic
correlations we predict are around $10-20\%$ of the currently
measured signal, somewhat smaller than the $1 \sigma$ errors on the 
measurements.

Since the area of the sky surveyed for weak lensing is increasing 
rapidly, the intrinisic correlation may become detectable from these
deep and wide surveys
in the future. In any case, it seems to be worth bearing in
mind that there could be this sort of contamination.
In particular, one possible way of extracting more information from
 lensing which has received attention is the use of 
photometric redshift information, to break down the background galaxy
distribution into a number of ``screens''. This would enable
tomography to be carried out (e.g., Hu 1999). We have seen however that 
the intrinsic correlation may be quite large for these narrow
redshift bins, so that it might become comparable to the weak lensing 
signal (Fig. \ref{shearobs}).

 Of course, the extra information available
in the form of photometric redshifts is likely to be very useful for deciding
whether there is an intrinsic component, and if it exists, to separate it from 
the lensing signal. For example the cross-correlation of ellipticities
(or co-variance of the shear) between different redshift bins can be compared
to the correlation within bins, with only the later responding to
intrinisic correlations. Something along these lines has
already been carried out by KWL, albeit with two colour bands which both
give wide redshift distributions (but with one deeper than the other).
These authors find a higher shear signal for the deeper redshift sample,
 which is consistent with lensing, but in the wrong direction for
intrinsic correlations. For the cross-correlation between samples, they 
do find slightly anomalous results, however.

Another way of trying to measure any intrinsic ellipticity correlations
would be to stick to the local universe, and  to measure the
three dimensional correlation functions (\S3) from a redshift survey.
If there really is a signal like that plotted in Fig. \ref{xie},
then this could be measurable from a relatively small survey (by todays
standards), with a few thousand galaxies. Even without redshifts, one
might expect to find a measurable intrinsic signal 
from a relatively nearby angular sample of galaxies, like 
the APM survey (Maddox \etal 1990), or Sloan Digital Sky Survey
(Gunn \& Weinberg 1995).

If there are in fact some measurable correlations between 
real galaxy ellipticities,
then this can be understood in the framework of structure formation by
gravitational instability, with the ellipticities being linked to 
the angular momenta of galaxies, which are in turn set up by tidal torques
from the shear in the initial density field (e.g., 
Peebles 1969,
Barnes \& Efstathiou 1987,
Catelan and Theuns 1996a,b).
This may explain why the ellipticity correlation functions we
measure have similar functional forms to those caused by weak-lensing: both
are responding to a cosmic shear field. Detection of
correlated ellipticities, if they exist, may be useful
for the study of galaxy formation (e.g., Sugerman \etal 2000), or
even cosmology (Lee \& Pen 2000).

It is also likely that the signal due to the intrinsic
correlation will give qualitatively and measurably different results
from weak lensing for some statistics we have not considered here. For example,
the probability distribution of the lensing convergence 
is predicted to have a measurable skewness, something which can be used to
determine $\Omega$ (Bernardeau \etal 1997). Measurements of this
parameter from our simulated surveys by L. Van Waerbeke 
(private communication) yield a null result, the convergence pdf being
consistent with a Gaussian distribution. The intrinisic correlations do
not therefore appear to interfere with our ability to do cosmology in this
way, and should not act as more than an additional source of noise
(albeit correlated) when reconstructed mass maps are made. 

On the simulation side, one important issue is the fact that our results
have apparently not converged with resolution. Although we find that 
the  higher resolution of two simulations gives more intrinsic correlations,
it is possible that given even higher resolution, things will begin
to go the other way. Clearly this needs to be tested somehow in the future.
Also, perhaps most important of all, we have assumed a very simple
relationship between projected halo ellipticities and 
projected galaxy  ellipticities. It is possible that  adding
gas dynamics and star formation to simulations will result in their
being no significant correlation between the two. The tests which we
have carried out which have most bearing on this are the use of two sets of  
different friends of friends groups, which respond to ellipticities
either of the whole halo, or just the dense central region. As we find results
for the two which are very similar, this is at least some
evidence that the intrinsic correlation may be fundamental.

\bigskip
As this paper was being completed, we became aware of
 similar work by Heavens
\etal (2000). These authors use the angular momentum of Nbody halos (also
from Virgo simulations, but only at the lower of the two resolutions)
to predict the intrinsic correlation of spiral galaxy ellipticities. 
They reach final results which are broadly similar (although they find
much more noise), and also conclude that 
while these effects are likely to be minor for present surveys, they may
become important with small redshift widths (in their case for shallower
surveys). We also became aware of analytic work on a similar theme
by Catelan \etal (in preparation), and Mackey and White (in preparation).

\bigskip
\acknowledgments
We thank Ludo Van Waerbeke,
Martin White, Volker Springel,
Marc Kamionkowski,
and Jordi Miralda-Escud\'{e} for useful discussions,
and Nick Kaiser for help in interpreting some observational results.
We also thank George Efstathiou for supplying us
with the FOF groupfinder and the Virgo consortium for making their simulation
data public. RACC acknowledges 
support from NASA Astrophysical Theory Grant NAG5-3820.

\end{document}